\documentstyle[12pt]{article}
\renewcommand{\tiny}{\rm}
\newcommand{\be}{\begin{equation}}
\newcommand{\ee}{\end{equation}}
\newcommand{\bq}{\begin{eqnarray}}
\newcommand{\eq}{\end{eqnarray}}
\newcounter{saveeqn}
\newcounter{App} 
\newcommand{\appa}{%
\stepcounter{App}%
\setcounter{saveeqn}{\value{equation}}%
\setcounter{equation}{0}%
\renewcommand{\theequation}{\Alph{App}\arabic{equation}} }
\newcommand{\appende}{%
\setcounter{equation}{\value{saveeqn}}%
\renewcommand{\theequation}{\arabic{equation}}  }
\begin{document}

\pagestyle{empty}
\hskip 3.0truein{DFUPG 69/92 December, 1992}
\vskip 0.2truein
\begin{center}
{\large \bf SU($N$) ANTIFERROMAGNETS AND THE PHASE STRUCTURE OF QED IN
THE STRONG COUPLING LIMIT}\\
\vspace{1 cm}
{\large M. C. Diamantini}$^{\mbox{\tiny 1}}$ {\large and P.
Sodano}\footnote{This work is supported in part by a grant from the
M.U.R.S.T..  M.C.D.  acknowledges the hospitality of the Physics
Department at the University of British Columbia where some of this
work was completed.}\\
\vspace{0.4 cm}
{Dipartimento di Fisica and Sezione I.N.F.N., Universit\'a di Perugia\\
Via A. Pascoli, 06100 Perugia, Italy}\\

\vspace{0.5 cm}
{\large E. Langmann}\footnote{Work supported in part by the ``Fonds zur
F\"orderung der wissenschaftlichen Forschung'' of Austria under contract Nr.
J0789-PHY.} {\large and G. W. Semenoff}\footnote{This work is
supported in part by the
Natural Sciences and Engineering Research Council of Canada.  G.S.
acknowledges the hospitality of the Physics Department of the
University of Perugia and I.N.F.N., Sezione di Perugia
where part of this work was completed.}\\
\vspace{0.4 cm}
{ Department of Physics, University of British Columbia\\
Vancouver, B.C., Canada V6T 1Z1}
\end{center}
\newpage
\begin{center}
{\bf Abstract}
\end{center}
We examine the strong coupling limit of both compact and non-compact
quantum electrodynamics (QED) on a lattice with staggered Fermions.
We show that every SU($N_{\tiny L}$) quantum antiferromagnet with
spins in a particular fundamental representation of the SU($N_{\tiny
L}$) Lie algebra and with nearest neighbor couplings on a bipartite
lattice is exactly equivalent to the infinite coupling limit of
lattice QED with the number of flavors of electrons related to
$N_{\tiny L}$ and the dimension of spacetime, $D+1$.  There are $N_{\tiny L}$
2-component Fermions in $D=1$, $2N_{\tiny L}$ 2-component Fermions in
$D=2$ and $2N_{\tiny L}$ 4-component Fermions in $D=3$.  We find that,
for both compact and non-compact QED, when $N_{\tiny L}$ is odd the
ground state of the strong coupling limit breaks chiral symmetry in
any dimensions and for any $N_{\tiny L}$ and the condensate is an
isoscalar mass operator.  When $N_{\tiny L}$ is even, chiral symmetry
is broken if $D\geq 2$ and if $N_{\tiny L}$ is small enough and the
order parameter is an isovector mass operator.  We also find the exact
ground state of the lattice Coulomb gas as well as a variety of
related lattice statistical systems with long--ranged interactions.
\newpage
\pagestyle{plain}
\setcounter{page}{1}

\section{Introduction}

It is reasonably well established that, as the bare coupling constant
of massless quantum electrodynamics (QED) is increased, there is a
phase transition which breaks chiral symmetry and generates an
electron mass.  The mechanism is similar to that in the
Nambu-Jona-Lasinio model \cite{NJL} where chiral symmetry breaking
occurs when the four--Fermion interaction is sufficiently strong and
attractive that a bound state and the resulting symmetry breaking
condensate forms. In the case of QED it is the Coulomb attraction of
electrons and positrons which, as the electric charge is increased,
gets strong enough to form a condensate. In QED, this phase transition
has been seen in both three and four spacetime dimensions using
numerical as well as approximate analytic techniques.  In four
dimensions, numerical simulations of compact \cite{kogut1987} and
noncompact
\cite{kogut et al 1988} lattice
QED with dynamical Fermions indicate presence of a phase transition as
the bare electric coupling is increased.  For compact QED the
transition is first order and for non--compact QED it appears to be of
second order.

For non-compact QED this phase transition is also found in the
continuum using approximate analytical techniques such as the solution
of Schwinger-Dyson equations in the quenched ladder approximation
\cite{fk,fomin et.al.,miransky,llm,llb1,llb2}.  The critical
behavior has the additional interesting feature that certain
perturbatively non-renormalizable operators such as four--Fermion
operators can become relevant there \cite{llb1,llb2}.

In three spacetime dimensions, a similar behavior is found in the
large $N$ expansion \cite{Applequist,Nash} where the inverse of the
number of Fermion flavors, $1/N$, plays the role of coupling constant.
Both analytical techniques \cite{Applequist,Nash} and numerical
simulations \cite{kogutnew} find a critical value of $N$, above which
QED is chirally symmetric and below which the chiral symmetry is
broken.

Furthermore, there are some recent analytic proofs that the strong
coupling limit of QED breaks chiral symmetry.  Salmhofer and Seiler
\cite{salm} showed using a Euclidean spacetime approach and staggered
Fermions that in four dimensions, four-flavor QED (as well as some
other U$(N_{\tiny C})$ gauge theories with $N_{\tiny C}\leq 4$) has a
chiral symmetry breaking ground state when the electric charge is
infinite.  Subsequently, using staggered Fermions in the Hamiltonian
approach it was shown \cite{s,ls,dsls} that the strong coupling limit
of QED has a chiral symmetry breaking ground state in any spacetime
dimension greater than two and when there are a specific number of
Fermion flavors (two four-component in 2+1 and four four-component
Fermions in 3+1 dimensions, also for other U$(N_{\tiny C})$ gauge
theories with any $N_{\tiny C}$).  This is an elaboration of previous
arguments for chiral symmetry breaking \cite{kogut1}
using strong coupling Hamiltonian methods (for a review, see
\cite{Kogutreview}).  Also, it is partially motivated by an
interesting previous work of Smit \cite{smit} where he uses naive and
Wilson Fermions (and also finds a particular kind of mapping to an
SU($N$) antiferromagent) to analyze chiral symmetry breaking in QED.

If the chiral symmetry breaking phase transition in four dimensions is
of second order, as it seems to be in non--compact QED
\cite{kogut et al 1988},
it provides a nonperturbative zero of the beta-function for
the renormalized coupling constant.  Furthermore, if the critical
behavior there differs from mean field theory, the resulting
ultraviolet fixed point would allow QED, in the limit of infinite
cutoff, to avoid the Landau pole, or Moscow zero
\cite{Landau,LandauPomeranchuk,EFradkin} which otherwise
renders it trivial.  It is an interesting and nontrivial question
whether, via this mechanism, QED could be an example of a nontrivial
field theory which exists in four dimensions.

The existence of the strong-coupling phase transition has long been
advocated by Miransky \cite{fomin et.al.,miransky}.
He visualized the mechanism
for chiral symmetry breaking as a ``collapse'' of the
electron-positron wave--function, similar to the behavior of the
supercritical hydrogen atom with bare proton charge Z$>$137.  He
argued that there are two ways to screen supercritical charges.  In
the case of a supercritical nucleus, the high electric field produces
electron-positron pairs, ejects the positrons and absorbs the
electrons to screen its charge.  In the case of supercritical electron
and positron charge, he argued that the pair production is suppressed
by Fermion mass, so a system can stabilize itself by increasing the
electron and positron masses; thus the tendency to break chiral
symmetry.  Miransky studied continuum QED in the quenched
approximation, using the ladder Schwinger-Dyson equations to sum the
planar photon-exchange graphs.  In this approximation, there is a line
of critical points, beginning at bare coupling $e^2=0$ and ending at
$e_c^2=\pi/3$ where the theory breaks chiral symmetry dynamically and
has an interesting continuum limit.  A behavior very similar to this
was found in lattice simulations using quenched
Fermions\cite{barth,hkd}.  At least some of its qualitative features
are expected to survive the presence of Fermion loops in realistic
QED.

In this Paper we shall examine Miransky's collapse phenomenon in a more
physical context by studying a lattice version of QED which is  similar
to a condensed matter system.  We do this by arguing that the lattice
approximation to QED with staggered Fermions in any number of dimensions,
$D>1$, resembles
a condensed matter system of lattice electrons in an external magnetic
field and with a half-filled band (this argument was also given in
\cite{s,ls,dsls}).  In that picture, the breaking of
chiral symmetry and the analog of the collapse phenomenon is the formation
of either charge or isospin density waves and the resulting reduction
of the lattice translation symmetry from translations by one site to
translations by two sites.  It occurs when the
exchange interaction of the electrons, which is attractive,  dominates the
tendency of the direct Coulomb
interaction and the kinetic energy to delocalize charge,
giving an instability to the formation of commensurate
charge density waves.  This forms a gap in the Fermion spectrum and a
particular mass operator obtains a vacuum expectation value.
This gives an intuitive picture of how strong attractive interactions
in a field theory can form a coherent structure.  Here, the commensurate
density waves in the condensed matter system correspond
to a modulation of the vacuum charge or isospin density at
the ultraviolet cutoff wavelength in the field theory.

We are interested in massless quantum electrodynamics with action
\bq
S=\int d^{D+1}x\left(-\frac{\Lambda^{3-D}}{4e^2}F_{\mu\nu}F^{\mu\nu}
+\sum_{a=1}^{N_F}\bar\psi^a\gamma^\mu(i\partial_\mu+A_\mu)\psi^a\right)
\label{contqed}
\eq
where $\Lambda$ is the ultraviolet cutoff, $e$ is the dimensionless
electric charge and there are $N_F$ flavors of $2^{[(D+1)/2]}$--dimensional
Dirac spinors. (Here, $[(D+1)/2]$ is the largest integer less than or
equal to $(D+1)/2$.)  In even
dimensions (when D+1 is even), the flavor symmetry is
$SU_R(N_F)\times SU_L(N_F)$.
In odd dimensions there is no chirality and the flavor symmetry
is $SU(N_F)$.  What is usually referred to as chiral symmetry there is
a subgroup of the flavor group.
We shall use a lattice regularization of (\ref{contqed})
and study the limit as $e^2\rightarrow\infty$.  We shall use $N_L$ flavors
of staggered fermions.  In the naive continuum limit, this gives $N_F=N_L~$
2--component fermions in D=1, $N_F=2N_L~$ 2--component fermions in D=2 and
$N_F=
2N_L~$ 4--component fermions in D=3.
Though the lattice theory reduces to (\ref{contqed}) in
the naive continuum limit, the lattice regularization breaks part
of the flavor symmetry.  All of the operators which are not symmetric
are irrelevant and vanish as the lattice spacing is taken to zero.
In D=3, the $SU_R(N_F)\times SU_L(N_F)$ symmetry
is reduced to $SU(N_F/2)$  (in D=3, $N_L=N_F/2$)
and translation by one site in the
3 lattice directions.  These discrete transformations
correspond to discrete chiral transformations
in the continuum theory.  In D=2, the lattice has $SU(N_F/2)$
(in D=2, $N_L=N_F/2$)
symmetry and two
discrete (``chiral'') symmetries.  In D=1, the $SU_L(N_F)\times SU_R(N_F)$
symmetry is reduced to $SU(N_F)$  (in D=1, $N_L=N_F$)
and a discrete chiral transformation.  In
each case, the discrete chiral symmetry is enough to forbid Fermion mass
and it is the spontaneous breaking of this symmetry which we shall examine
and which we shall call ``chiral symmetry breaking''.  We shall also
consider the possibility of spontaneous
breaking of the $SU(N_F)$ flavor symmetry.  In
the continuum the $SU(N_F)$  corresponds to a vector--like symmetry
subgroup of the full flavor group.
Since the lattice and continuum theories have different symmetries, the
spectrum of Goldstone bosons, etc. would be different in the
two cases.  In the following, we
shall not address the source of these differences but will
define QED by its lattice regularization and discuss the realization of the
symmetries of that theory only.

In this Paper we shall prove that, in the strong coupling limit,
lattice QED with $N_{\tiny L}$ lattice flavors of staggered
Fermions is {\bf
exactly equivalent} to an SU$(N_{\tiny L})$ quantum antiferromagnet
where the spins are in a particular fundemental representation of the
SU$(N_{\tiny L})$ Lie algebra. Furthermore, mass operators of QED
correspond to staggered charge and isospin density operators in the
antiferromagnet.  Thus, the formation of charge-density waves
corresponds to chiral symmetry breaking and the
dynamical generation of an iso--scalar Fermion mass whereas
N\'eel order of the antiferromagnet
corresponds to dynamical generation of Fermion mass with an
iso--vector condensate.    As
a result of this correspondence
we can use some of the properties of the quantum
antiferromagnets to deduce features of strongly coupled QED.

We find that, in the infinite coupling limit, the properties of the
electronic ground state of compact and non-compact QED are identical.
The fact that compact QED confines and non-compact QED (at least in high
enough dimensions) does not confine affects only the properties of the gauge
field wavefunctions and the elementary excitations.
\vskip 0.2truein

\noindent
We find that the case of even $N_{\tiny L}$ and odd $N_{\tiny L}$ are
very different:
\vskip 0.1truein

When $N_{\tiny L}$ is odd, the vacuum energy in the strong coupling
limit is proprotional to $e^2$, the square of the electromagnetic
coupling constant.  We also find that chiral symmetry is broken in the
strong coupling limit for all odd $N_{\tiny L}\geq 1$ and for all
spacetime dimensions $D+1\geq2$.  The mass operator which obtains a
nonzero expectation value is a Lorentz and flavor Lie algebra scalar.
There are also mass operators which are Lorentz scalars and which
transform nontrivially under the flavor group which could get an
expectation value and break the flavor symmetry spontaneously if
$N_{\tiny L}$ is small enough.  We expect (but do not prove) that when
$N_{\tiny L}$ increases to some critical value there is a phase
transition to a disordered phase.

In contrast, when $N_{\tiny L}$ is even we find that the vacuum energy
in the strong coupling limit is of order 1.  We find that chiral
symmetry is broken when the spacetime dimension\footnote{1+1
dimensions is a special case which we will discuss later
(Section 4).}
$D+1\geq 3$ and when $N_{\tiny L}$ is small enough.  The mass operator
which gets an expectation value is a Lorentz scalar and transforms
nontrivially under the flavor SU$(N_{\tiny L})$.  Thus, flavor
symmetry is spontaneously broken.  As in the case of odd $N_{\tiny
L}$, we expect that there is an upper critical $N_{\tiny L}$ where
there is a transition to a disordered phase.

In the course of our analysis, we find the exact ground state of the
generalized classical Coulomb gas model in $D$--dimensions with Hamiltonian
\bq
H_{\rm coul}=\frac{e^2}{2}\sum_{<x,y>}\rho(x)g(x-y)\rho(y)
\eq
where the variable $\rho(x)$ lives at the sites of a square lattice with
spacing one (with coordinates $(x_1,\ldots,x_D)$ and $x_i$ are integers)
and takes on values
\bq
-\frac{N_{\tiny L}}{2},-\frac{N_{\tiny L}}{2}+1,\ldots,\frac{N_{\tiny L}}{2}+1,
\frac{N_{\tiny L}}{2}
{}~~N_{\tiny L}~{\rm an~odd~integer}
\eq
and the interaction is long-ranged
\bq
g(x-y)\sim\vert x-y\vert^{2-D}~~,~~{\rm as}~\vert x-y\vert\rightarrow\infty
\eq
In any space dimension, $D\geq1$ we prove that there are two degenerate
ground states which have the
Wigner lattice configurations
\bq
\rho(x)=\pm {1\over2}(-1)^{\sum_{i=1}^D x_i}
\eq
When $N_{\tiny L}=1$ this is a long-ranged Ising model which is known to have
antiferromagnetic order, even in one dimension.  It is also in agreement
with a known result about the Ising model in two dimensions \cite{halsey}.

The difference between even and odd $N_{\tiny L}$ can be seen to
arise from a certain frustration encountered when, with odd $N_{\tiny L}$,
one simultaneously
imposes the conditions on lattice QED which should lead to charge conjugation
invariance, translation invariance  and Lorentz invariance of the continuum
limit.  This frustration is absent when $N_{\tiny L}$ is even.  This is not an
anomaly in the conventional sense of the axial anomaly or a discrete anomaly
encountered in the quantization of gauge theories, as no symmetries are
incompatible, but it is nevertheless an interesting analog of the anomaly
phenomenon.

Note that in 3 spacetime dimensions the difference between even and odd
$N_{\tiny L}$ is the difference between an even
and odd number of 4-component Fermions.  Our result that for an odd number
of 4-component Fermions the chiral symmetry is always broken for large
coupling seems to contradict the continuum analysis in \cite{Applequist,Nash}.
We do not fully understand the reason for this, but speculate that
it is related to the lattice regularization.  Their model is very
similar to ours in that they effectively work in the strong coupling
limit when they replace the ultraviolet regularization which comes from
having a Maxwell term in the QED action by a large momentum cutoff.  In our
case we have a lattice cut-off and the result should be very similar.
Note that we agree with them when $N_{\tiny L}$ is
even and there is an even number of 4-component Fermions.  In that case, we
are also in qualitative agreement with recent numerical work \cite{kogutnew}
which, since it uses Euclidean staggered Fermions and there is a further
Fermion doubling due to discretization of time, can only consider the
case where there is an even number of 4-component Fermions. The anomalous
behavior that we find with odd numbers of 4-component Fermions exposes a
difficulty with treating the flavor number $N$ as a continuous parameter.

A hint as to why 2+1--dimensional Fermions should come in
four-component units appears when we formulate compact QED in the continuum
using the SO(3) Georgi-Glashow
model with spontaneous breaking of the global symmetry, $SO(3)
\rightarrow U(1)$ and the limit of large Higgs mass \cite{poly}.  We begin
with the model
\bq
{\cal L}= \frac{1}{4e^2}\sum_{a=1}^3(F^a_{\mu\nu})^2+\frac{1}{2}\sum_{a=1}^3
(D\phi)^a\cdot(D\phi)^a+\frac{\lambda}{4}
((\sum_{a=1}^3\phi^a\phi^a)-v^2)^2
\eq
The spectrum contains a massless photon which in the $\lambda\rightarrow
\infty$ and $v^2\rightarrow\infty$ is the only light excitation.
Since the U(1) gauge group is a subgroup of SO(3), it is compact.  We
wish to make 2+1-dimensional electrodynamics by coupling this model to
Fermions.  It is known that, if the resulting theory is to preserve
parity and gauge invariance simultaneously, we must use an even number
of two-component Fermions which, in the minimal case are also SU(2)
doublets \cite{niemi}
\bq
{\cal L}_F=\bar\psi\left(i\gamma\cdot\nabla+\gamma\cdot A+
\gamma^5(m+g\phi\cdot\sigma)\right)\psi
\eq
Here, the Fermions are four-component and have a parity invariant mass
and Higgs coupling.  Each SU(2) doublet contains two electrons (which
can be defined so that both components of the doublet have the same
sign of electric charge), thus the basic Fermion for compact QED has
eight components.  The maximal chiral symmetry is obtained in the
massless case with vanishing Higgs coupling.  It is possible, by
suitable choice of mass and Higgs coupling i.e. $m=\pm g<\phi>$, (and
also reduction of the chiral symmetry) to make four of the Fermions
heavy, leaving four massless components.  This is consistent with
parity and gauge invariance.  It is also interesting to note that
staggered Fermions on a Euclidean lattice, where time is also
discretized, produce eight component continuum Fermions.

In the lattice gauge theory, the coupling of gauge
fields to the electron field does not distinguish between compact and non-
compact QED.  Therefore, the above considerations for the Fermion content
should apply to both cases.

Similarly, in four dimensions it is known that the Fermion
multiplicity obtained by lattice regularization is the one which is
compatible with the axial anomaly \cite{karstensmit,nielsen}.  An
alternative way to see this constraint is to again form compact QED using
the Georgi-Glashow model.  Then, because of Witten's SU(2) anomaly
\cite{witten}, we are required to use at least two (and generally an even
number of) 2-component Weyl Fermions which are at least SU(2) doublets.
This seems to indicate that the minimal Fermion would have eight components.
This is the minimal number that we find using staggered Fermions in the
Hamiltonian approach (see Appendix B).  It is interesting that in Euclidean
staggered Fermions produce sixteen components, and therefore in our
terminology always lead to the case with even $N_{\tiny L}$.

A different context where a correspondence between spin systems and gauge
theory appears is in the study of strongly correlated electron systems in
condensed matter physics.  An important issue
there is the correspondence of spin systems such as the quantum
Heisenberg antiferromagnet (which is equivalent to the strong
coupling limit of the Hubbard model at 1/2 filling)
and certain limits of
various kinds of lattice gauge theories.
Gauge theory-like states can be obtained as an approximate low energy
theory in the mean field approximation.  An example is the ``flux
phases'' suggested in \cite{weigman,anderson,Affleck 1988}
 and also various dimerized phases \cite{suth}
which are disordered and under some conditions compete with the N\'eel
phase, particularly in 2+1 dimensions (for a review, see \cite{fradbk}).
Affleck and
Marston \cite{Affleck 1988} showed how to get the flux phase from mean
field theory.  It is a locally stable but not global minimum of the
free energy of an antiferromagnetic spin system and could presumably
be stabilized by adding certain operators to the Hamiltonian.  The
low-energy limit in this phase resembles strongly coupled 2+1--dimensional
lattice QED
with four species of massless 2-component electrons (because of the two spin
states of the lattice electrons, this is the case
$N_{\tiny L}=2$ in 2+1-dimensions).

The picture that we shall advocate in this paper is that there should
be a phase transition between the flux phase and the N\'eel ordered
phase of the antiferromagnet.  This phase transition is governed by
the strength of the effective electromagnetic coupling constant.  The
Heisenberg antiferromagnet, which is known to have a N\'eel ordered
ground state, is obtained in the limit of infinite electric coupling
constant.  For weaker couplings, the system can be in the flux phase
where the electrons are massless.  As the coupling is increased, the
N\'eel state is recovered by formation of a commensurate spin density
wave\footnote{This is a commensurate charge density wave for each spin
state whose phase is such that the condensate has zero electric
charge.}, which corresponds to spontaneous chiral symmetry breaking
and the generation of iso--vector Fermion mass in the effective QED.
(Spin in the condensed matter system corresponds to iso--spin in
effective QED.)

In Section II we discuss the formulation of QED on a lattice.
We give a detailed discussion of discrete
symmetries and also of gauge fixing which is necessary to make non--compact
QED well--defined.  In Section III we discuss the
strong coupling limit and show how quantum antiferromagnets are
obtained in the strong coupling limit for both compact and non-compact
QED.  We discuss the properties of the electron ground state for both cases
of $N_L$ even and odd.
We also discuss the implications of the mapping between antiferromagnets
and strong coupling gauge theories for the symmetries of the ground state.
Section IV is devoted to concluding remarks.

We review some of our notation in Appendix A and the essential
features of the staggered Fermion formalism, with emphasis on those
aspects which are important for our arguments, in Appendix B.  Appendix C
is devoted to a review of the Fermion formulation of spin systems.
\newpage

\section{\bf QED on a Lattice}

In this Section we shall set the Hamiltonian formalism of Abelian
gauge fields on a lattice.  For the most part, this formalism
can be found in some of the classic reviews of lattice gauge theory,
\cite{Kogutreview} for example.  A novel feature of the present
Section is a careful treatment of the normal ordering of the charge
operator and a discussion of the ensuing discrete symmetries.  This
normal ordering turns out to be important if the continuum limit is to have
the correct behavior under $C$, $P$ and $T$ transformations.  It will also
be important in our later solution of the
strong coupling limit.

\subsection{\bf Hamiltonian and Gauge Constraint}

We shall discretize space as a cubic lattice and, in order to use the
Hamiltonian formalism, time is left continuous.  We use units in which
the lattice spacing, the speed of light and Planck's constant are all
equal to one.  (See Appendix A for a summary of our notation and
Appendix B for a review of staggered Fermions.)  Lattice gauge fields
are introduced through the link operators
\bq
U_i(x)\equiv e^{iA_i(x)}
\label{3.1}
\eq
which correspond to the link $[x,i]$ of the lattice.  Electric fields
propagate on links of the lattice and the electric field operator
$E_i(x)$ associated with the link $[x,i]$ is the canonical conjugate
of the gauge field
\bq
\bigl[ A_i(x), E_j(y)\bigr]=i\delta_{ij}\delta(x-y)
\label{3.2}
\eq
The gauge field and electric field operators obey the relations
\bq
A_{-i}(x)
= -A_i(x-\hat i)~~~,~~E_{-i}(x)=-E_i(x-\hat i)
\eq
The Hamiltonian (of non--compact QED) is
\bq
H_{\rm
NC}=\sum_{[x,i]}{e^2\over2}E^2_i(x)+\sum_{[x,i,j]}{1\over2e^2}B^2[x,i,j]
{}~~~~~~~~~~~~~~~~~~~~\nonumber\\
+\sum_{[x,i]}\left(
t_{[x,i]}\psi^{a\dagger}(x+i)e^{iA_i(x)}\psi^a(x)+{\rm h.c.}\right)
\label{3.3}
\eq
where the second term contains a sum over plaquettes and the magnetic
field is defined as the curvature of the gauge field at the plaquette
$[x,i,j]$,
\bq
B[x,i,j]=A_i(x)+A_j(x+\hat i)+A_{-i}(x+\hat i+\hat j)+A_{-j}(x+\hat
j)\nonumber\\ =A_i(x)-A_j(x)+A_j(x+\hat i)-A_i(x+\hat j)
\label{3.4}
\eq
As is discussed in Appendix B, the hopping parameter $t_{[x,i]}$
contains phases which produce a background magnetic flux $\pi$ per
plaquette.  In the weak coupling continuum limit, the magnitude of
$t_{[x,i]}$ is one, $\vert t\vert^2\equiv\vert t_{[x,i]}\vert^2=1$ in
order that the speed of the free photon and free electron fields are
equal, i.e. so that the low frequency dispersion relations for both
the photon and electron have the same speed of light.  However, in
order to obtain a relativistic continuum limit in general it is
necessary to make $\vert t\vert$ a function of $e^2$.  We shall find
that in the limit where $e^2$ is large, the speed of light is
proportional to $\vert t\vert/e$ and it is necessary that $\vert
t\vert \sim e$ as $e^2\rightarrow
\infty$.

The Hamiltonian we have written in (\ref{3.3}) is appropriate to
non-compact QED.  If we wish to study compact QED we must make the
Hamiltonian symmetric under the field translation
\bq
A_i(x)\rightarrow A_i(x)+2\pi
\label{LGT}
\eq
for any $[x,i]$.  This is accomplished by replacing the second term in
(\ref{3.3}) by a periodic function of the magnetic flux, so that (for
compact QED)
\bq
H_{\rm C}=\sum_{[x,i]}{e^2\over2}E_i(x)^2+
\sum_{[x,i,j]}{2\over e^2}\sin^2\left(B[x,i,j]/2\right)\nonumber\\
+\sum_{[x,i]}\left(t_{[x,i]}\psi^{a\dagger}(x+i)e^{iA_i(x)}\psi^a(x)+
{\rm h.c.}\right)
\label{3.3a}
\eq
Both (\ref{3.3}) and (\ref{3.3a}) reduce to the standard Hamiltonian
of QED in the naive weak coupling continuum limit.  Away from that
limit the behavior of the dynamical systems described by the two
Hamiltonians can differ significantly.  For example, in the strong
coupling limit compact QED is a confining \cite{poly} theory whereas
non-compact QED in not confining.  Also, the phase transition seen in
numerical simulation of the two theories differ.  In the compact case
the phase transition associated with chiral symmetry breaking is
generally of first order \cite{kogut1987} whereas it is second order
for the non--compact case \cite{kogut et al 1988}.
The source of some of these differences generally
have to do with the symmetry (\ref{LGT}).  In (\ref{3.3}) and
(\ref{3.3a}) we have introduced $N_{\tiny L}$ flavors of lattice
Fermions labelled by the index $a=1,\ldots,N_{\tiny L}$.

In both compact and non--compact QED, the Hamiltonian is supplemented with
the constraint of gauge
invariance.  The gauge transformations of the dynamical variables,
\bq
\Lambda:&&A_i(x)\rightarrow A_i(x)+\nabla_i\Lambda(x)
\nonumber\\
\Lambda:&&E_i(x)\rightarrow E_i(x)
\nonumber\\
\Lambda:&&\psi^a(x)\rightarrow e^{i\Lambda(x)}\psi^a(x)
\nonumber\\
\Lambda:&&\psi^{a\dagger}(x)\rightarrow\psi^{a\dagger}(x)e^{-i\Lambda(x)}
\label{3.5}
\eq
are generated by the operator
\bq
{\cal G}_\Lambda\equiv\sum_x
\left(-\nabla_i\Lambda(x) E_i(x)+\Lambda(x)
\left(\psi^{a\dagger}(x)\psi^a(x)-N_{\tiny L}/2\right)\right)
\label{3.6}
\eq
The local generator of gauge transformations where $\Lambda$ has
compact support is
\bq
{\partial{\cal G}_\Lambda\over\partial\Lambda(x)}\equiv{\cal
G}(x)=\hat\nabla\cdot E(x)+\psi^{a\dagger}(x)\psi^a(x)-N_{\tiny L}/2
\label{gauss}
\eq
Both (\ref{3.6}) and (\ref{gauss}) commute with the Hamiltonians in
(\ref{3.3}) and (\ref{3.3a}).

In (\ref{3.6}) and (\ref{gauss}) we have subtracted the constant
$N_{\tiny L}/2$ from the charge density operator in order to make the
gauge generator odd under the usual charge conjugation transformation
\bq
\xi:&&A_i(x)~\rightarrow ~-A_i(x)
\nonumber\\
\xi:&&E_i(x)~\rightarrow ~-E_i(x)
\nonumber\\
\xi:&&\psi^a(x)~\rightarrow ~(-1)^{\sum_{k=1}^{D}x_k}\psi^{a\dagger}(x)
\nonumber\\
\xi:&&\psi^{a\dagger}(x)~\rightarrow~ (-1)^{\sum_{k=1}^{D}x_k}\psi^a(x)
\label{xi}
\eq
In fact, the Fermionic charge term in (\ref{3.6}) can be put in the
manifestly odd form ${1\over2}[\psi^{a\dagger}(x),\psi^a(x)]$ Of
course, charge conjugation symmetry of the lattice theory is necessary
to obtain charge conjugation of the continuum theory.  We shall see
later that, particularly at strong coupling, the subtraction term in
(\ref{3.6}) plays an important role.  It seems to have been ignored in
previous literature ( for example, see
\cite{kogut1,Kogutreview}).  Its presence is particularly important
when $N_{\tiny L}$ is odd since the charge operator has no zero
eigenvalues in that case (the eigenvalues of $\psi^{\dagger}\psi$ are
integers).

Chiral symmetry is related to translation invariance by one site.  The
Hamiltonians (\ref{3.3}) and (\ref{3.3a}) are invariant under the
transformations
\bq
\chi_j:&&A_i(x)~\rightarrow~ A_i(x+\hat j)
\nonumber\\
\chi_j:&&E_i(x)~\rightarrow~ E_i(x+\hat j)
\nonumber\\
\chi_j:&&\psi^a(x)~\rightarrow ~(-1)^{\sum_{k=j+1}^{D}x_k}
\psi^a(x+\hat j)
\nonumber\\
\chi_j:&&\psi^{a\dagger}(x)~\rightarrow ~(-1)^{\sum_{k=j+1}^{D}x_k}
\psi^{a\dagger}(x+\hat j)
\label{chi}
\eq
for $j=1,\cdots, D$.

In the following we shall use the charge conjugation symmetry which is
a combination of these two transformations:
\bq
C\equiv\xi\chi_1
\label{C}
\eq
This is necessary if the mass operators which we define in Appendix B is to be
invariant under charge conjugation symmetry.  Also, we shall see that
the strong coupling ground state is invariant under $C$ but not under
either $\xi$ or $\chi_j$ by themselves.

The dynamical problem of Hamiltonian lattice gauge theory is to find
the eigenstates of the Hamiltonian operator (\ref{3.3}) or
(\ref{3.3a}) and out of those eigenstates to find the ones which are
gauge invariant, i.e. which obey the physical state condition (or, the
``Gauss' law'' constraint)
\bq
{\cal G}(x)\vert \Psi_{\rm phys.}>=0
\label{3.7}
\eq
Note that the gauge constraint and physical state condition are the
same for both compact and non--compact QED.  In the case of compact
QED there is the additional symmetry (\ref{LGT}) which, being a large
gauge symmetry, can be represented projectively.  In fact, when we
later work in the Schr\"odinger picture we shall require that the
quantum states which are functions of a configuration of the gauge
field transform as
\bq
\vert A_i(x)+2\pi n>=\exp\left(in\theta[x,i]\right)\vert A_i(x)>
\label{THETA}
\eq
There is a separate parameter $\theta[x,i]$ for each link of the
lattice.  These parameters originate in a way similar to the
theta--angle in ordinary QCD.  The symmetry (\ref{THETA}) together
with the commutator (\ref{3.2}) imply that the spectrum of the
electric field operator has eigenvalues which are separated by
integers and offset by $\theta$:
\bq
{\rm spectrum}[E_i(x)]=\{\theta[x,i]+{\rm integers}\}
\label{specE}
\eq

The Hamiltonian and gauge constraints can be obtained from the gauge
invariant Lagrangian
\bq
L=\sum_x\psi^{a\dagger}(x)(i\partial_t-A_0(x))\psi^a(x)+\sum_{[x,i]}
E_i(x)\dot A_i(x)\nonumber\\ +\sum_{[x,i]} E_i(x)\nabla_i
A_0(x)+\sum_xA_0(x)N_{\tiny L}/2-H
\label{3.8}
\eq
where the temporal component of the gauge field has been introduced to
enforce the gauge constraint and the time derivative terms give the
correct symplectic structure.  Note that, in order to get Lorentz
invariance of the Fermion spectrum in the weak coupling (naive) continuum
limit, we require half-filling of the Fermionic states, i.e. that the total
charge defined by
$$
\sum_x \left( \psi^{a\dagger}(x)\psi^a(x)-N_{\tiny L}/2\right)
$$
has zero vacuum expectation value.

Here we have considered massless QED.  As well as the gauge invariance
and charge conjugation invariance discussed above, the Hamiltonian is
symmetric under the discrete chiral transformations (\ref{chi}) which
on the lattice corresponds to symmetry under translation by one site.
In later Sections, we shall consider the possibility of spontaneous
breaking of this symmetry.

\subsection{\bf Gauge fixing and quantization}

We shall quantize the gauge fields in the Sch\"odinger picture.  The
quantum states are functions of the link operators which are
represented by the classical variables $A_i(x)$ and the electric field
operators are derivatives
\bq
E_i(x)\equiv -i{\partial\over\partial A_i(x)}
\label{3.9}
\eq
We shall also consider the usual Fock representation of the Fermion
anticommutator.  The empty vacuum is the cyclic vector
\bq
\psi^a(x)\vert 0>=0~~~\forall a,x
\label{3.10}
\eq
and Fermions occupying lattice sites are created by $\psi^{a\dagger}(x)$.

\subsubsection{Compact QED}

In compact QED the spectrum of the gauge generator is discrete and a
state which obeys the physical state condition can be normalized, thus
implying that there is no need for additional gauge conditions.  The
basis wave-functions for compact QED (in the basis where the Fermions
density and eletric field operators are diagonal)
are $\Psi[n(x)]\Phi[A]$ with the Fermion states
\bq
\Psi[n(x)] = \prod_x \prod_{a=1}^{N_{\tiny L}}
(\psi^{a\dagger}(x))^{n_a(x)} \vert 0>
\label{fermionstates}
\eq
labeled by vectors $n(x)=(n_1(x),\cdots,n_{N_{\tiny L}}(x))$ with
$n_a(x)=0$ or $1$, and the photon states
\bq
\Psi[A]=\exp\left(i\sum e_i(x)A_i(x)\right)
\label{states}
\eq
where the eigenvalues $e_i(x)$ of the electric field operator are in
spectrum$[E_i(x)]$ (\ref{specE}).  Furthermore the states of the
photon field are normalized using the inner product
\bq
<\Phi_1[A],\Phi_2[A]>=
\prod_{[x,i]}\int_0^{2\pi}{dA_i(x)\over2\pi}\Phi_1^{\dagger}[A]\Phi_2[A]
\eq
and the Fermion states have conventional inner product given by
$<0\vert 0>=1$ and the canonical anticommutator relations of the
fermion field operators. The physical state condition (\ref{3.7})
gives the additional restriction that
\bq
\hat\nabla_ie_i(x)=-\rho(x)= -\sum_x\sum_a(n_a(x)-1/2)
\eq
where $\rho(x)$ is the charge density (i.e. the eigenvalue of
$\psi^{a\dagger}(x)\psi^a(x)-N_{\tiny L}/2$).  Pictorially, we can
think of this as containing lines of electric flux joining sites whose
charges are non-zero and also closed loops of electric flux. In the
strong coupling limit, the Hamiltonian is diagonal in the basis
(\ref{states}). This gives a natural explanation of confinement in
compact QED in the strong coupling region.  If we add a
particle--antiparticle pair to a state in (\ref{states}) it must be
accompanied by at least a single line of electric flux.  The energy of
such a line of flux is proportional to its length.  Therefore the
electron-positron interaction grows linearly with distance and is
confining.  This is in contrast to the situation in non-compact QED
where the electric flux is not quantized.  In that case, a
particle--antiparticle pair can have many lines with arbitrarily small
flux. The energy of the field is minimized by the usual Coulomb dipole
configuration.  In high enough dimensions this is not a confining
interaction.

\subsubsection{Non--Compact QED}

In contrast to the case of compact QED, in non--compact QED the
generator ${\cal G}(x)$ (\ref{gauss}) of gauge transformations has a
continuous spectrum.  In order to obtain a normalizable ground state
it is therefore necessary to fix the additional gauge freedom.

In order to separate the gauge orbits of the photon field, we shall
need to define the transverse and longitudinal components of the gauge
fields.  The transverse projection operator is
\bq
T_{ij}=\delta_{ij}-{\nabla_i\hat\nabla_j\over\nabla\cdot\hat\nabla}
\label{3.11}
\eq
and the longitudinal projection operator is
\bq
L_{ij}={\nabla_i\hat\nabla_j\over\nabla\cdot\hat\nabla}
\label{3.12}
\eq
They have the usual property of projection operators,
\bq
T^2=T
{}~~~
L^2=L
{}~~~
TL=0=LT
\label{3.13}
\eq
and also,
\bq
1=T+L
\label{3.14}
\eq
The transverse and longitudinal parts of the electric and gauge fields are
obtained by
\bq
A^{\tiny T}_i(x)\equiv T_{ij}A_j(x)
{}~~~
A^L_i(x)\equiv L_{ij}A_j(x)
\label{3.15}
\eq
\bq
E^{\tiny T}_i(x)\equiv T_{ij}E_j(x)
{}~~~
E^{\tiny L}_i(x)\equiv L_{ij}E_j(x)
\label{3.20}
\eq
Note that there is an ambiguity in the precise definition of the
projection operators $T$ and $L$ on a finite lattice due to the zero
mode of the lattice laplacian. We have fixed this so that constant
fields, $A_i(x)=a_i=const.$, are purely transverse, $a_i=a_i^T$ (see
Appendix A).

In order to quantize, it is necessary to solve the
gauge constraint
\bq
{\cal G}(x) = \hat\nabla\cdot E(x) + \rho(x) \sim 0
\label{Gausslaw}
\eq
with
\bq
\rho(x) \equiv \psi^{a\dagger}(x)\psi^a(x) - N_{\tiny L}/2
\label{rho}
\eq
This is most easily accomplished by gauge fixing.  The procedure
\cite{dirac} is to find a gauge fixing condition which has a nonzero
commutator with the gauge constraint.  An example is the Coulomb gauge
condition
\bq
\chi(x)=\hat\nabla\cdot A(x) \sim 0
\label{3.21}
\eq
The commutator of the gauge condition with the
gauge generator is
\bq
\bigl[ \chi(x), {\cal G}(y)\bigr]= -i\nabla\cdot\hat\nabla\delta(x-y)
\label{3.22}
\eq
which is a non-degenerate matrix.  We can then solve the two constraints
by eliminating the longitudinal parts of the gauge and electric fields,
\bq
A^{\tiny L}_i(x)=0
\label{3.23}
\eq
\bq
E^{\tiny L}_i(x)= -\nabla_i {1\over-\nabla\cdot\hat\nabla}\rho(x)
\equiv
-\sum_y \nabla_i(x\vert {1\over-\nabla\cdot\hat\nabla}\vert y)\rho(y)
\label{3.24}
\eq
The remaining degrees of freedom obey canonical commutation relations
which are derived from Dirac brackets.  These brackets project the
canonical Poisson brackets onto the constrained phase space.  Given a
set of constraints, $\xi_A$ with a non-degenerate Poisson bracket,
$\det
\left\{\xi_A,\xi_B\right\}_{PB}\neq 0$, Dirac brackets for dynamical
variables are obtained from Poisson brackets as
\bq
\left\{ P,Q\right\}_{DB}=\left\{ P,Q\right\}_{PB}-\left\{ P,\xi_A
\right\}_{PB}\left\{\xi_A,\xi_B\right\}_{PB}^{-1}\left\{\xi_B,Q\right\}_{PB}
\label{3.26}
\eq
In the present case, the brackets of the remaining variables are not
modified.  The commutator for the transverse photon and electric
fields is
\bq
\bigl[ A_i^{\tiny T}(x), E^{\tiny T}_j(y)\bigr]=i T_{ij}\delta(x-y)
\label{3.27}
\eq
The Hamiltonian depends only on the transverse photon and electric
field and on the charged Fermion fields,
\bq
H=\sum_{[x,i]}{e^2\over2} (E^{\tiny
T}_i(x))^2+\sum_{[x,i]}{1\over2e^2}A^{\tiny T}_i(x)(-\nabla\cdot
\hat\nabla)A^{\tiny T}_i(x)\nonumber\\
+\sum_{x}{e^2\over2}\rho(x){1\over-\nabla\cdot\hat\nabla}\rho(x)
\nonumber\\
+\sum_{[x,i]}t_{[x,i]}\left(\psi^{a\dagger}(x+i)e^{iA^{\tiny T}_i(x)}
\psi^a(x)+ {\rm h.c.}\right)
\label{3.28}
\eq
The Coulomb interaction has appeared as a result of the solution of
Gauss' law.

This gauge fixing, which we have done following Dirac's procedure
\cite{dirac}, can always be implemented by a canonical transformation.
Gauss' law is solved by taking the ansatz for the physical states
\bq
\vert{\rm phys}>=\exp\left(i\sum_{[x,i]} A_i(x)\cdot\nabla_i
\frac{1}{-\nabla\cdot\hat\nabla} \rho(x)\right)
\vert A^T>\equiv U\vert A^T>
\label{phys}
\eq
The exponential operator in (\ref{phys}) generates the unitary
transformation
\bq
\tilde A_i(x)=UA_i(x)U^{\dagger}=A_i(x)
\nonumber\\
\tilde E_i(x)=UE_i(x)U^{\dagger}=
E_i(x)- \nabla_i \frac{1}{\hat\nabla\cdot\nabla}\rho(x)
\nonumber\\
\tilde\psi(x)=U\psi(x)U^{\dagger}= \exp\left( \frac{i}{\hat\nabla\cdot
\nabla}\hat\nabla\cdot A(x)\right)\psi(x)
\nonumber\\
\tilde\psi^{\dagger}(x)=U\psi^{\dagger}(x)U^{\dagger}=\psi^{\dagger}(x)\exp
\left( -\frac{i}{\hat\nabla\cdot\nabla}\hat\nabla\cdot A(x)\right)
\label{canon}
\eq
Note that the transformation of $\psi$ and $\psi^{\dagger}$ removes
the longitudinal part of the gauge field from the covariant hopping
term,
\bq
\tilde\psi^{\dagger}(x+i)e^{i\tilde A_i(x)}\tilde\psi(x)=\psi^{\dagger}(x+i
)e^{iA_i^{\tiny T}(x)}\psi(x)
\eq
Substituting the canonically transformed fields into the Hamiltonian
yields the Hamiltonian (\ref{3.28}) which is decoupled from $A^{\tiny
L}_i(x)$ together with the canoncially transformed Gauss' law which
now states that $E^{\tiny L}=0$.  The longitudinal parts $E^{\tiny L}$
and $A^{\tiny L}$ can now be dropped from the phase space and the
resulting quantum theory is the same as that obtained from Dirac's
procedure for solving the constraints.


For $D=1$ (1+1 dimensional spacetime), (\ref{3.28}) is the Hamiltonian
of the lattice Schwinger model. It is worth pointing out that in that
case, $A^T$ and $E^T$ just comprise one quantum mechanical degree of
freedom,
$$
A^T=\sum_x A(x), \quad E^T= \sum_x E(x)
$$
This corresponds to the fact that the only physical degree of freedom
of the photon field on a 1 dimensional compact space is the Wilson
loop (U(1) holonomy) $\exp(i\sum_x A(x))$.

\section{\bf Strong Coupling Expansion}

Although the results are very similar, the implementation of the
strong coupling expansion is somewhat different in the two cases of
compact and non--compact QED.  We shall treat the two cases
separately.

\subsection{Non--Compact QED}

The conventional strong coupling, $e^2\rightarrow\infty$ limit is
difficult to implement for noncompact QED since the leading terms in
$e^2$ in the Hamiltonian have a continuum spectrum.  The alternative,
but related procedure is the hopping parameter expansion, i.e. an
expansion in the parameter $\vert t\vert$ in equation (\ref{3.28}).
The terms in this expansion are very similar to a strong coupling
expansion in that they contain inverse powers of $e^2$.

We begin by separating the Hamiltonian into two parts, a leading order
part
\bq
H_0=\sum_{[x,i]}{e^2\over2}(E^{\tiny T}_i(x))^2+
\sum_{[x,i]}{1\over2e^2}A^{\tiny T}_i(x) (-\nabla\cdot
\hat\nabla)A^{\tiny T}_i(x)
+\nonumber\\
\sum_{x,y}
{e^2\over 2}\rho(x)(x\vert{1\over-\nabla\cdot\hat\nabla}\vert y)\rho(y)
\label{ho}
\eq
whose ground state we shall attempt to find exactly and a
next-to-leading order part
\bq
H_1=\sum_{[x,i]}t_{[x,i]}\left( \psi^{a\dagger}(x+i)e^{iA^{\tiny
T}_i(x)} \psi^a(x)+ {\rm h.c.} \right)
\eq
which we treat as a perturbation.

We first examine the structure of the ground state of $H_0$.  First of
all, it is a direct sum of the free transverse photon Hamiltonian and
the Coulomb interaction which depends only on the Fermion operators.
The wave-function therefore factorizes into a wavefunction for the
free photon ground state and a wavefunction for the ground state of
the four-Fermion operator in (\ref{ho}).  In the Schr\"odinger
picture, the photon ground state is the Gaussian
\bq
\Phi_{\rm photon}[A]=
{1\over C}\exp\left\{-{1\over2e^2}\sum_x A_i^{\tiny
T}(x)\sqrt{-\nabla\cdot\hat\nabla} A_i^{\tiny T}(x)\right\}
\eq
($C$ the normalization constant) and the photon contribution to the
ground state energy is just the ground state energy of the free photon
theory and is of order zero in $e^2$ as well as $\vert t\vert$.

The nature of the ground state of the four-Fermion part depends on the
number of Fermion flavors, $N_{\tiny L}$.  In particular it is quite
different when $N_{\tiny L}$ is even or odd and we shall treat these
two cases separately.

\subsubsection{\bf $N_{\tiny L}$ Even}

If $N_{\tiny L}$ is an even number, the ground state of the operator
\bq
H_{\rm coul}=\sum_{x,y}\frac{e^2}{2}\rho(x)(x\vert
\frac{1}{-\nabla\cdot\hat\nabla}
\vert y)\rho(y)
\label{hcoul}
\eq
is the state $\vert g.s.>$ where
\bq
\sum_{a=1}^{N_L}\psi^{^a\dagger}(x)\psi^a(x)
{}~\vert g.s.>= N_{\tiny L}/2~\vert g.s.>
\label{cnstr}
\eq
i.e. with every site of the lattice half-occupied. It is easy to see
that this is the case by noting that $H_{\rm coul}$ is a
non--negative operator and that the states with zero charge density
are zero eigenvalues of $H_{\rm coul}$.

This ground state is degenerate.  At each site the quantum state is given by
\bq
\prod_{i=1}^{N_{\tiny L}/2}\psi^{a_i\dagger}\vert 0>
\eq
Since this quantity is antisymmetric in the indices
$a_1,\ldots,a_{N_{\tiny L}/2}$ it takes on any orientation of the
vector in the representation of the flavor symmetry group SU($N_{\tiny
L}$) with Young Tableau made of one column with $N_{\tiny L}/2$ boxes
($m=N_{\tiny L}/2$ in Fig.\ 1).
\vskip 4.0truein
\noindent
{\bf Figure 1:}The representation of $SU(N_L)$ at each site when $N_L$ is
even.

\vskip 0.25truein

As in ref. \cite{s,ls,dsls} we
observe that the degeneracy must be resolved by diagonalizing
perturbations in the hopping parameter expansion.  The first order
perturbations vanish.  The first non--trivial order is second order,
\bq
\delta_2= -<g.s.\vert H_1{1\over H_0-E_0}H_1\vert g.s.>
\eq
This matrix element can be evaluated by noting that $H_1$ creates an
eigenstate of $H_0$ different from the ground states with additional
energy
\bq
\Delta E=\frac{e^2}{2}+\frac{e^2}{2}(D)\nabla_1
(x\vert\frac{1}{-\nabla\cdot\hat\nabla}\vert x)~=~e^2
\eq
Diagonalizing the matrix of second order perturbations is equivalent
to finding the spectrum of the effective Hamiltonian
\bq
H_{\rm eff}={2\vert t\vert^2\over e^2}\sum_{[x,i]}\psi^{b\dagger}(x+\hat
i)\psi^b(x)\psi^{a\dagger}(x)\psi^a(x+\hat i)
\nonumber\\
=\frac{2\vert t\vert^2}{e^2}\sum_{[x,i]}J_{ab}(x)J_{ba}(x+\hat i)
\label{heff}
\eq
where the operators
$J_{ab}(x)=\psi^{a\dagger}(x)\psi^b(x)-\frac{1}{2}\delta_{ab}$, are
the generators of the U($N_{\tiny L}$) given in equation
(\ref{generators}) of Appendix C and obeying the Lie algebra in
equation (\ref{algebra}).

The constraint (\ref{cnstr}) on the total occupation number of each
site, $$
\rho(x)=\sum_{a=1}^{N_L} J_{aa}(x) \sim 0
$$ reduces to SU($N_{\tiny L}$) (see Appendix C) and projects onto the
irreducible representation given by the Young Tableau in Fig.\ 1.
(This is one of the fundamental representations of SU($N_{\tiny L}$).)
Furthermore, (\ref{heff}) is just the Hamiltonian of the SU($N_{\tiny
L}$) antiferromagnet in that representation.

It is straightforward to see that the higher orders in the hopping
parameter expansion also have higher orders of $1/e^2$.  In fact, if
we consider the following limit,
\bq
e^2\rightarrow\infty~~~,~~\vert t\vert^2\rightarrow\infty
\nonumber\\
\vert t\vert^2/e^2=~{\rm constant}
\label{limit}
\eq
all higher order perturbative contributions to both the wavefunction
and the energy vanish.  Thus, in this limit, QED is {\bf exactly}
equivalent to an SU($N_{\tiny L}$) antiferromagnet.  That (\ref{limit}) is the
correct limit to take can be seen from the fact that, if the
antiferromagnet in (\ref{heff}) is in an ordered state, the speed of the
spin-waves, which are the gapless low-energy excitations is proportional to
$\vert t\vert/e$.  They have linear dispersion relation $\omega(k)\sim\vert
k\vert$ and play the role of massless goldstone bosons for broken flavor
symmetry.  Their speed should be equal to the speed of light, which is one
in our units.  This implies that $\vert t\vert/e$ should be adjusted so
that the spin-wave spectrum is relativistic, $\omega(k)=\vert k\vert$.
Hence the limit in (\ref{limit}).

When $N_{\tiny L}=2$, this model is the quantum Heisenberg
antiferromagnet in the $j=1/2$ representation.  It is known to have a
N\'eel ordered ground state in $D\geq3$ \cite{lieb} and there is good
numerical evidence that it has N\'eel order in D=2.  The
antiferromagnetic order parameter is the mass operator
\bq
\vec\Sigma = \sum_x (-1)^{\sum_{i=1}^D x_i}\psi^{\dagger}(x)\vec\sigma\psi(x)
\eq
which obtains a vacuum expectation value in the infinite volume limit.
Thus, when $N_L=2$
the strong coupling limit breaks chiral symmetry and generates
electron mass.  It is interesting that in this case there is an
iso--vector condensate.  In the strong coupling limit this seems
unavoidable.  The only way to get an iso--scalar condensate is with a
charge density wave.  However such a configuration always has infinite
coulomb energy compared to an electric charge
neutral but isospin carrying condensate.

The low energy excitations of this systems (with energies of order
$\vert t\vert^2/e^2$ are spin waves.  All other excitations have
energies which go to infinity in the limit (\ref{limit}).  The spin waves
are the pions which are the scalar Goldstone bosons arising from spontaneous
breaking of the vector flavor symmetry $SU(2)\rightarrow U(1)$.

For large $N_{\tiny L}$ there is some evidence that the SU$(N_{\tiny L})$
antiferromagnet in these specific representations has a disordered
ground state \cite{read2}.  Particularly in 2+1-dimensions it is known that
for infinite $N_L$ the ground state is disordered \cite{Affleck 1988}
Although it is beyond the scope of this
paper, it would be intersting to
investigate the $N_{\tiny L}$ dependence of
the ground state further.  We shall comment on this in Section IV.

\subsubsection{\bf $N_{\tiny L}$ Odd}

When $N_{\tiny L}$ is odd the charge density operator $\rho(x)$
(\ref{rho}) which enters the Coulomb Hamiltonian (\ref{hcoul}) has no
zero eigenvalues.  Therefore the Coulomb energy of the ground state is
necessarily of order $e^2$ for large $e^2$.

Since the Coulomb Hamiltonian commutes with the charge density
operator, $\rho(x)$, they can be diagonalized simultaneously.  The ground
state of the Coulomb Hamiltonian should therefore
also be an eigenstate of $\rho(x)$.
Therefore, to find the spectrum of (\ref{hcoul}) we consider all
states which are also eigenfunctions of the local density, i.e. where at a
given site $x$,
\bq
\rho(x)=-N_{\tiny L}/2,-N_{\tiny L}/2+1,\dots,N_{\tiny L}/2
\eq
with the constraint of global neutrality
\bq
\sum_x\rho(x)=0
\nonumber
\eq

For convenience, we consider the system on a finite spatial lattice
$V=V_R$ with periodic boundary conditions (see Appendix A). Then the
momenta $k$ appearing in the Fourier transform are discrete.  We consider
the Coulomb Hamiltonian in momentum space
\bq
H_{\rm coul}\equiv \frac{e^2}{2}\sum_{x,y}\rho(x)(x\vert\frac{1}{
-\hat\nabla\cdot\nabla}\vert y)\rho(x)
\nonumber\\
=\frac{e^2}{2}\frac{1}{|V|} \sum_k \frac{1}{4\sum_{i=1}^D \sin^2
(k_i/2)}\vert \tilde\rho(k)\vert^2
\label{coulomb}
\eq
where $|V|$ is the total number of spatial lattice sites and $\tilde\rho (k
)$ is the fourier transform of the charge operator $\rho(x)$.  Since $0\leq
\sin^2(k_i/2)\leq 1$ we can derive the lower bound for the Coulomb
energy as $$ H_{\rm coul}\geq
\frac{e^2}{2}\frac{1}{|V|}\sum_k \frac{1}{4D}\vert\tilde\rho(k)\vert^2
=\frac{e^2}{8D}\sum_x\rho(x)^2\geq\frac{|V| e^2}{32D} $$
(the latter estimate follows from $\rho(x)^2 \geq 1/4$) This bound is
saturated by the charge distributions $$ \tilde\rho_0(k)=
\pm\frac{|V|}{2}\delta_{\vec k,\vec \pi} $$ where
$\vec\pi=(\pi,\ldots,\pi)$ is the vector for which
$\sum_{i=1}^D\sin^2(k_i/2) $ in the denominator of (\ref{coulomb})
takes its maximum value.  These are allowed configurations of the
charge density, $$ \rho_0(x)=\pm\frac{1}{2}(-1)^{\sum_{i=1}^D x_i}$$
which give the ground state configurations of the Coulomb system.  The
electric field in these ground states is easily deduced from Gauss' law,
$$ E_i^0(x)= \pm\frac{1}{4D}(-1)^{\sum_{i=1}^D x_i} $$ These configurations
break chiral symmetry in that they are not invariant under the
transformation $\chi_1$ in (\ref{chi}) but they are symmetric under
charge conjugation $C$ defined in (\ref{C}).

The ground state energy per lattice site is
\bq
E_{0(\rm coul)}/|V|=\frac{e^2}{32D}
\eq
Note that it is of order $e^2$.  This is in contrast to the ground state
energy when the number of lattice Fermion flavors is even, which is of
order $\vert t\vert^2/e^2\sim 1$.

The ground states that we have found are highly degenerate in that only
the number of Fermions at each site is fixed.  Their quantum state can
still take up any orientation in the vector space which carries the
representation of the flavor SU($N_{\tiny L}$) given by the Young
Tableaux in Fig.\ 2.
\vskip 4.0truein
\noindent
{\bf Figure 2:} Representation of $SU(N_L)$ on each site of the even
sublattice $A$ and the odd sublattice $B$ when $N_L$ is odd.

\vskip 0.25truein

We have divided the lattice into two sublattices: $A$ is all points where
$\sum_i x_i$ is even and $B$ where $\sum_i x_i$ is odd.  Then, the
differing occupation numbers on sites on each sublattice yield different
representations of $SU(N_L)$.

Again, this degeneracy must be resolved by diagonalizing the
perturbations, which are non-zero in second order and the problem is
equivalent to diagonalizing the antiferromagnet Hamiltonian
(\ref{heff}), this time with the representations depicted in Fig.\ 2.
Also, in the limit (\ref{limit}) this correspondence is {\bf exact}.

The strong coupling ground states that we find when $N_{\tiny L}$ is
odd contains a charge density wave.  The staggered charge density
operator has expectation values
\bq
\frac{1}{|V|}<\sum_x(-1)^{\sum_{k=1}^{D}x_k}\psi^{\dagger}(x)\psi(x)>=
\pm 1/2
\eq
This condensate is an isoscalar
and we have shown that it must always occur in all dimensions.  When
$N_{\tiny L}>1$ mass operators with certain generators of SU($N_{\tiny
L}$) could have expectation values if the ground state has
antiferromagnetic order.  However, unlike the case of even $N_{\tiny
L}$, the antiferromagnetic order is not required in order to have
chiral symmetry breaking.

The ground state we find breaks chiral symmetry.  This is a true
dynamical symmetry breaking since, in infinite volume, the ground
states which are related by a chiral transformation are never mixed in
any order of strong coupling perturbation theory.  Furthermore, there
are no local operators which couple them.

We conclude that the strong coupling ground state breaks chiral symmetry
for any odd $N_{\tiny L}$ and in any dimensions.  As in the case of even
$N_L$ there is also the possibility (and for small $N_L$ the likelyhood)
that the $SU(N_L)$ antiferromagnet we obtain here is in a N\'eel state and
the flavor symmetry is also broken.  We shall not pursue this possibility
here but refer the reader to the literature \cite{read2}.

\subsection{Compact QED}

The difference between compact and non-compact QED resides in the
quantization of the gauge fields.  In all cases the Fermionic state is
identical in the two cases.  In compact QED the eigenstates of the
electric field operator are normalizable and can be used for the
ground state.  In this case we separate the Hamiltonian into three
terms,
\bq
H_0=\sum_{[x,i]}\frac{e^2}{2}E_i^2(x)\nonumber\\
H_1=\sum_{[x,i]}t_{[x,i]}\left(\psi^{\dagger}(x)e^{iA_i(x)}\psi(x)+{\rm
h.c.}\right)
\nonumber\\
H_2=\sum_{[x,i,j]}\frac{2}{e^2}\sin^2\left(B[x,i,j]/2\right)
\eq
In the strong coupling limit it is necessary to solve Gauss' law
(\ref{3.7}) for the electric fields and the charge distribution in
such a way as to minimize $H_0$.

When $N_{\tiny L}$ is even, the charge operator has zero eigenvalues
and Gauss' law has the solution where $E_i(x)=0$, which is an obvious
minimum of the $H_0$, and there are $N_{\tiny L}/2$ fermions on each
site.  This is similar to the situation in non--compact QED when
$N_{\tiny L}$ is even.  Also, the degeneracy of this state must be
resolved in the same way, resulting in the effective Hamiltonian
(\ref{heff}) which describes the SU($N_{\tiny L}$) antiferromagnet in
the representation with Young Tableau having one column with $N_{\tiny
L}/2$ boxes shown in fig. 1.
Again, we expect that this system has N\'eel order in $D\geq2$ if
$N_{\tiny L}$ is small enough and the chiral symmetry of
electrodynamics is broken, along with the SU($N_{\tiny L}$) flavor
symmetry.

When $N_{\tiny L}$ is odd, since the charge density operator has no
non-zero eigenvalues, it is impossible to find a zero eigenstate of
the Gauss' law constraint operator without some electric field.  The
problem which we must solve is to minimize the energy functional $
\sum E^2$ subject to the constraint $\hat\nabla\cdot E=-\rho$ where at
each site $\rho$ has one of the values $$ -\frac{N_{\tiny
L}}{2},-\frac{N_{\tiny L}}{2}+1,\ldots,\frac{N_{\tiny
L}}{2}~~~N_{\tiny L}~{\rm an~odd~integer} $$ It is straightforward to
show that the charge distribution and electric field which one obtains
is identical to those in the case of non--compact QED with odd
$N_{\tiny L}$, $$
\rho_0(x)={1\over2}(-1)^{\sum_{i=1}^D x_i}~~~E_i(x)=
\frac{1}{4D}(-1)^{\sum_{i=1}^D x_i}
$$ The ground state degeneracy is again resolved by diagonalizing
perturbations and, again the true ground state of the strong coupling
limit is the ground state of the effective Hamiltonian (\ref{heff})
when the SU($N_{\tiny L}$) spins take on the configurations in
Fig.2.

Notice that in the ground state, the electric fields are not integers, but
on each link, the spectrum of the electric field operator is $1/4D$+
integers. The ``theta angles'' 1/4D survive all orders in strong coupling
perturbation thery.

\section{\bf Remarks}

In this paper we have analyzed the possibility of chiral symmetry breaking
in the strong coupling limit of quantum electrodynamics using the
Hamiltonian picture and a lattice regularization.  We chose to use
staggered Fermions because they give the closest analog to interesting
condensed matter physics systems.  Also, unlike Wilson Fermions which, in
the Hamiltonian picture,  have
no chiral symmetry at all, they have a discrete chiral invariance which
forbids Fermion mass and it is sensible to ask questions about dynamical
mass generation.

In 1+1 dimensions, staggered Fermions give $N_{\tiny L}$ species of 2-
component Dirac Fermions.  When $N_{\tiny L}=1$ we obtain the
Schwinger model with a lattice regularization.  Also, in this case, we
have found that the chiral symmetry is broken dynamically.  Of course,
due to the staggered Fermion regularization there is no continuous
chiral symmetry, which is as it should be since it should be
impossible to regularize the Schwinger model so that there is
simultaneous continuous chiral and gauge symmetry.  However, to match the
solution of the continuum Schwinger model, the Fermion should obtain
mass.  This indeed happens in our strong coupling limit by spontaneous
symmetry breaking.  (Although we disagree with some aspects of the
formalism, we agree with the results of reference \cite{kogut1} on this
point.)

This result should not be confined to strong
coupling, but should persist for all coupling, i.e. the critical
coupling in $D=1$ should be at $e^2=0$.  We conjecture that this sort of
symmetry breaking for small $e^2$ is a manifestation of the Peierls
instability --- the tendency of a one dimensional Fermi gas to from a
gap at the Fermi surface.  This happens with any infinitesimal
interaction.

In fact, this must also happen for the case where $N_{\tiny L}$ is even.
Then, there cannot be any spin order in 1 dimension.  However, anomalies
break the isoscalar chiral symmetry in the continuum theory and should also
do so here.  This means that there should be a dynamical generation of
charge density wave which would be driven by the Peierls insability.  It
also implies that for $N_{\tiny L}=2$ for example,
the ground state in the strong coupling limit would not be a
Heisenberg antiferromagnet, but would be alternating empty site and site
with two electrons in a spin singlet state.  This state, even though it has
large coulomb energy, avoids the infrared divergences of gapless Fermions.

In higher dimensions, $D\geq2$, it would be interesting to explore the
possibility of phase transitions between different symmetry breaking
patterns for the SU($N_{\tiny L}$) flavor symmetry as one varies
$N_{\tiny L}$.  There is already some work on this subject in the
condensed matter physics literature on SU($N_{\tiny L}$)
antiferromagnets \cite{read2}.  They analyze the SU($N_{\tiny L}$)
antiferromagnet which is similar to the strong coupling limit of an
U($N_{\tiny C}$) gauge theory (see \cite{ls,dsls} for details) which
is in the representation corresponding to a rectangular Young Tableau
with $N_{\tiny L}$ rows and $N_{\tiny C}$ columns.  They work in the
large $N_{\tiny C}$ limit and show that there is a phase transition
from the spin ordered N\'eel phase to a disordered phase when
$N_{\tiny L}\sim N_{\tiny C}$.  In our case $N_{\tiny C}=1$ so their
analysis is not accurate.  Nevertheless, we expect that there should
be a phase transition to a disordered phase as $N_{\tiny C}$ is
increased.  For odd $N_{\tiny L}$ the chiral symmetry is always broken
and the question we are asking is whether the flavor symmetry is also
broken.  For even $N_{\tiny L}$ possible phase transition is relevant
to both chiral and flavor symmetry breaking.

First of all, when $N_{\tiny C}=2$, we have the $j=1/2$ Heisenberg
antiferromagnet which is known to have an ordered ground state in
$D\geq3$ and is also very likely to have an ordered ground state in
$D= 2$.  Furthermore, when $N_{\tiny L}\rightarrow\infty$, the ground
state is known to be disordered in $D=2$ \cite{Affleck 1988} and is
likely also the case in $D=3$.  In between there $N_{\tiny L}=2$ and
$N_{\tiny L}=\infty$ there should be a phase transition.  It is
interesting to speculate that the order--disorder phase transition
which occurs as one increases $N_{\tiny L}$ in the SU($N_{\tiny L}$)
antiferromagnet is the same one that appears in the study of chiral
symmetry breaking in 2+1--dimensional QED in the continuum
\cite{Applequist,Nash} where they find that chiral symmetry is broken
only if the number of flavors is less than a certain critical value.
As we noted in the introduction, their work is effectively in the
strong coupling (large $e^2$) limit.

Our results also indicate that, besides the critical $N_{\tiny L}$,
for a fixed $N_{\tiny
L}$ which is small enough, there should also be a critical coupling
constant $e^2$ and, in fact, a critical line in the $N_{\tiny L}$--$e^2$
plane where there is a second order phase transition between a spin ordered
chiral symmetry breaking phase and a disordered  (and possibly chirally
symmetric phase).   We speculate that in 3+1--dimensions a similar
situation could occur.

\appendix
\appa
\section{Notation}

In this paper we consider the lattice regularization most suited to the
Hamiltonian formalism where time is continuous and space is a square
lattice with lattice spacing one.
We use a finite spatial lattice $V_R$ with lattice sites
\bq x\equiv (x_1,\ldots,x_{D}), \quad -R\leq x_i < R
\eq
were $R$ is a positive integer and $|V_R|= (2R)^D$ is the
total number of lattice sites. In the thermodynamic limit, $R\to
\infty$.

The lattice sites are connected by unit vectors
\bq
\hat 1=(1,0,\dots)
{}~~
\hat 2=(0,1,\dots)
{}~~
\dots
{}~~
\eq
and the oriented link between the lattice site $x$ and $x+\hat i$ is
denoted $[x,i]$.  The link oriented in the opposite direction is
denoted $-[x,i]$.  On the finite lattice $V_R$ we identify lattice
sites $x$ and $x+2R\hat i$. Then links obey the identity
\bq
[x,-i]=-[x-\hat i,i]
\eq
The boundary of the $[x,i]$ are the two points
\bq
\delta[x,i]=(x+\hat i) - x
\eq
Also, a plaquette with corners $x,x+\hat i,x+\hat i+\hat j,x+\hat j$
and with sides $[x,i],[x+\hat i,j],[x+\hat i+\hat j,-i],[x+\hat j,-j]$
is denoted as $[x,i,j]$ and has the boundary
\bq
\delta[x,i,j]= [x,i]-[x,j]+[x+\hat i,j]-[x+\hat j,i]
\eq
It also obeys the identities
\bq
[x,j,i]=-[x,i,j] ~~~
[x,-i,j]=-[x-\hat i,i,j]
\eq
It is also possible to introduce higher dimensional structures,
elementary cubes, etc.

We shall also introduce lattice derivative operators, the forward
difference operator
\bq
\nabla_i f(x)= f(x+\hat i)-f(x)
\eq
and the backward difference operator
\bq
\hat\nabla_i f(x)= f(x)-f(x-\hat i)
\eq
The lattice Laplacian is
\bq
\nabla\cdot\hat\nabla f(x)\equiv\sum_{i=1}^{D} \left(
f(x+\hat i)-2f(x)+f(x-\hat i)
\right)
\eq

The functions which we shall consider are functions from either the
lattice sites, links or plaquettes to the real numbers.  The Fourier
transform of a function on lattice sites is given by
\bq
\tilde f(k)= \sum_{x\in V_R} e^{ik\cdot x}f(x)
\label{fourier}
\eq
with $\tilde f$ a function on the reciprocal lattice (momentum space)
$\tilde V_R$ with lattice sites
\bq
k=(k_1,\ldots, k_D),\quad k_i=\frac{2\pi}{2R}\times{\rm integers}, \quad
-\pi< k_i\leq \pi
\eq
The inverse Fourier transform is
\bq
f(x)= \frac{1}{|V_R|}\sum_ke^{-ik\cdot x}\tilde f(k)
\eq
(note the the number of sites of the lattice $V_R$ and its reciprocal
lattice $\tilde V_R$ are equal). The periodic delta functions on $V_R$
and $\tilde V_R$ are given by
\bq
\delta(x)=\frac{1}{|V_R|}\sum_x e^{-ik\cdot x}
\eq
and
\bq
\delta(k)=\frac{1}{|V_R|}\sum_{k\in\tilde V_R}e^{ik\cdot x}
\eq
and the Parseval relation is
\bq
\sum_{x\in V_R} f^*(x)g(x) = \frac{1}{|V_R|}\sum_{k\in \tilde V_R}
\tilde f^*(-k)g(k)
\eq
In momentum space, the lattice Laplacian is diagonal,
\bq
\widetilde{(\nabla\cdot\hat\nabla f)}(k) =
-4\sum_{i=1}^D\sin^2(k_i/2)\tilde f(k)
\eq
{}From this it follows that its inverse
$\frac{1}{\nabla\cdot\hat\nabla}$ is unambigously defined only on
functions $f$ with $\tilde f(0)=\sum_{x\in V_R}f(x) =0$.  We can extend
its definition to all functions $f$ by setting
\bq
\widetilde{\frac{1}{-\nabla\cdot\hat\nabla}\tilde f}(0) = 0
\eq
In position space, the integral kernel of
$\frac{1}{-\nabla\cdot\hat\nabla}$ is just the Green function for the
Laplacian,
\bq
(x\vert{1\over-\nabla\cdot\hat\nabla}\vert y)=
\sum_{k\in \tilde V_R \notin \{0\}} e^{ik\cdot(x-y)}{1\over
4\sum_{i=1}^D\sin^2(k_i/2)}
\eq

For the infinite lattice $V_\infty$, the momentum space is
no longer a lattice but the Brillouin zone
\bq
\tilde V_\infty= \Omega_B \equiv\{k=(k_1,\ldots,k_D)|
k_i\mbox{ real }, -\pi<k_i\leq\pi\}
\eq
The inverse of the Fourier transform (\ref{fourier}) is then
\bq
f(x)= \int_{\Omega_B} {d^{D}k\over(2\pi)^{D}}e^{-ik\cdot x}\tilde
f(k)
\eq
corresponding to the delta function on the infinite lattice,
\bq
\delta(x)= \int_{\Omega_B} {d^{D}k\over(2\pi)^{D}}e^{ik\cdot x}
\eq

The Green function for the Laplacian on $V_\infty$ is given by
\bq
(x\vert{1\over-\nabla\cdot\hat\nabla}\vert
y)=\int_{\Omega_B}{d^{D}k\over (2\pi)^{D}}e^{ik\cdot(x-y)}{1\over
4\sum_i\sin^2k_i/2}
\eq
and is well defined in dimensions $D>2$.  In
two dimensions it is defined by one additional subtraction which
removes the logarithmic divergence in the integration.  In one
dimension it must be defined by solving the Laplace equation with a
source explicitly.

In the main text we appreviate $\sum_{x\in V_R}$ as $\sum_x$ and
similarly in momentum space.

\appende
\appa
\section{\bf Lattice Fermions}

Throughout this paper we shall use the staggered Fermion formalism
which was originally developed by Kogut and Susskind.  This formalism
is well known and the details can be found in the papers of Susskind
and collaborators \cite{susskind,kogutsusskind,kogut1,Kogutreview}
and Kluberg-Stern et.al.
\cite{Kluberg}.  Here we shall review the basic features and make some
observations which are necessary for our present discussion.  Some of
these observations have already been made in \cite{s} \cite{ls}
\cite{dsls}.

We shall use the staggered Fermion formalism since we believe that it
gives the closest possible analog to the lattice Fermions encountered
in condensed matter physics.  As a regularization of Fermions in
relativistic quantum field theory, this formalism has the disadvantage
that chiral symmetries are discrete, rather than continuous.  The
method should be regarded as adding some formally irrelevant operators
to the Hamiltonian.  These operators make the Hamiltonian local but
break the continuous chiral symmetry down to a discrete subgroup.
(Actually, there is a non-local chiral symmetry.  However, being
non-local it is not a useful symmetry in that, for example, it does
not imply the existence of Goldstone Bosons in the phase where it is
broken.)

Thus, we can really only address questions about discrete chiral
symmetry breaking.  This should be enough to tell us whether mass
generation, and in fact what sort of mass generation, is possible.

\subsection{\bf Review of Staggered Fermions}

The purpose of the staggered Fermion method is to minimize Fermion
doubling which always accompanies lattice Fermions.  Generally,
staggered Fermions are obtained by the spin-diagonalization method.
To implement this method, we begin with the naively latticized Dirac
Hamiltonian,
\bq
H_F={1\over2}\sum_{[x,j]}\left(
\psi^{\dagger}(x)i\alpha^j\nabla_j\psi(x)-
(\nabla_j\psi^{\dagger})(x)i\alpha^j\psi(x)\right)
\label{2.1}
\eq
where $\alpha^j$ are the $2^{[(D+1)/2]}$-dimensional Dirac $\alpha$-matrices.
(Here $[(D+1)/2]$ is the integer part of $(D+1)/2$.)  They are Hermitean,
$\alpha^{j\dagger}=\alpha^j$ and obey the Clifford algebra
\bq
\left\{ \alpha^i,\alpha^j\right\}=2\delta^{ij}
\label{2.3}
\eq
They are therefore unitary matrices, $\alpha^{i\dagger}\alpha^i=1$.
Using the properties of the difference operator, (\ref{2.1}) can be
presented in the form
\bq
H_F=-{i\over2}\sum_{[x,j]}\left( \psi^{\dagger}(x+\hat j)\alpha^j\psi(x)-
\psi^{\dagger}(x)\alpha^j\psi(x+\hat j)\right)
\label{2.2}
\eq
Since the Dirac matrices are unitary, the naive lattice Fermion
Hamiltonian in (\ref{2.2}) resembles a condensed matter Fermion
hopping problem with a background $U(2^{[(D+1)/2]})$ gauge field given by
the $\alpha$-matrices.  In any plaquette of the lattice, $[x,i,j]$,
this background field has curvature
\bq
\alpha^i\alpha^j\alpha^{i\dagger}\alpha^{j\dagger}=-1
\label{2.4}
\eq
The curvature resides in the U(1) subgroup of $U(2^{[(D+1)/2]})$ and has
exactly half of a U(1) flux quantum per plaquette.  This is true in
any dimensions. We observe that either 1/2 or zero flux quanta are
the only ones allowed by translation invariance and parity and time
reversal symmetries of the Hamiltonian.

Since the curvature if the $\alpha$-matrices is U(1)-valued, we should
be able to do a gauge transform which presents the matrices themselves
as U(1) valued gauge fields (i.e. diagonal).  A specific example of
such a gauge transform due to Kluberg-Stern et. al. \cite{Kluberg} is
\bq
\psi(x)\rightarrow (\alpha^1)^{x_1}(\alpha^2)^{x_2}
\dots(\alpha^{D})^{x_{D}}\psi(x)
\label{2.5}
\eq
Then
\bq
\psi^{\dagger}(x+\hat j)\alpha^j\psi(x)\rightarrow (-1)^{\sum_{k=1}^{j-1}x_k}
\psi^{\dagger}(x+\hat j)\psi(x)
\label{2.6}
\eq
The resulting Hamiltonian is
\bq
H_F=-{i\over2}\sum_{[x,j]}(-1)^{\sum_{k=1}^{j-1}x_k}
\left(\psi^{\dagger}(x+\hat j)
\psi(x)-\psi^{\dagger}(x)\psi(x+\hat j)\right)
\label{2.7}
\eq
This describes $2^{[(D+1)/2]}$ identical copies of Fermions with the
same Hamiltonian which must all give Fermions with the same spectrum
as the original Hamiltonian in (\ref{2.1}).  Staggered Fermions are
obtained by choosing one of these copies.  This reduces the Fermion
doubling by a factor of the dimension of the Dirac matrices,
$2^{[(D+1)/2]}$.

In the staggered Fermion method, we treat the components of the
original lattice Dirac Hamiltonian as flavors, rather than components
of the relativistic spinor necessary for Lorentz invariance.  The
spinor components now reside on adjacent lattice sites.  In this
method, the continuous chiral symmetry of the massless Hamiltonian,
under the transformation $\psi\rightarrow e^{i\gamma^5\theta}\psi$, is
lost. There is a discrete chiral symmetry, corresponding to
translations by one lattice site in any direction.  Explicitly,
\bq
\psi(x)\rightarrow (-1)^{\sum_{k=j+1}^{D}x_k}\psi(x+\hat j)
\label{2.8}
\eq
is a symmetry of the Hamiltonian (\ref{2.7}) and corresponds to a
discrete chiral transformation.

Mass operators correspond to staggered charge densities.  The operator
\bq
\Sigma = \sum_x (-1)^{\sum_{k=1}^{D}x_k}\psi^{\dagger}(x)\psi(x)
\label{2.9}
\eq
changes sign under the chiral transformations (\ref{2.8}) and
corresponds to a certain Dirac mass.

With staggered Fermions there is still a certain amount of Fermion
doubling.  The doubling can be counted by noting that the staggered
Fermion Hamiltonian (\ref{2.7}) is invariant under translations by two
lattice sites.  Therefore, a unit cell is a unit hypercube of the
lattice, containing $2^{D}$ sites and staggered Fermions correspond
to a $2^{D}$ component spinor.  The dimension of the Dirac matrices
is $d^{[(D+1)/2]}$.  Therefore the number of Dirac spinors we obtain is
$2^{D}/2^{[(D+1)/2]}$.  For lower dimensions the minimum number of
continuum flavors can be tabulated as
$$
\matrix{ d&~{\rm dim.~of~Dirac~matrices}~&~{\rm No.~of~flavors}\cr
         1+1&   2& 1\cr
         2+1&   2& 2\cr
         3+1&   4& 2\cr  }
\label{matrix}
$$
Only in 1+1-dimensions do we get a single species of Dirac Fermion.

\subsection{\bf Explicit Example in 3+1 Dimensions}


For simplicity in notation, the formulas here and in the following
subsection are given for an infinite spatial lattice $V_\infty$.

To see how to take the continuum limit explicitly, consider the case
of $d=3+1$.  There, we divide the lattice into eight sublattices and
label the spinor components as
\bq
\psi({\rm even},{\rm even},{\rm even})\equiv \psi_1
{}~~~\psi({\rm odd},{\rm even},{\rm odd})\equiv \psi_7
\label{2.11a}
\eq
\bq
\psi({\rm even},{\rm odd},{\rm even})\equiv \psi_6
{}~~~\psi({\rm even},{\rm even},{\rm odd})\equiv \psi_5
\label{2.11b}
\eq
\bq
\psi({\rm odd},{\rm odd},{\rm even})\equiv \psi_4
{}~~~\psi({\rm odd},{\rm even},{\rm odd})\equiv \psi_3
\label{2.11c}
\eq
\bq
\psi({\rm even},{\rm odd},{\rm odd})\equiv \psi_2
{}~~~\psi({\rm odd},{\rm odd},{\rm odd})\equiv \psi_8
\label{2.11d}
\eq
In terms of these spinors, the Hamiltonian (\ref{2.7}) can be written
as the matrix operator
\bq
H=\int_{\tilde\Omega_B}{d^{3}k}
\psi^{\dagger}(k)A^i\sin k_i\psi(k)
\label{2.12}
\eq
where
\bq
\tilde \Omega_B=\{ k_i: -\pi/2<k_i\leq\pi/2\}
\eq
is the Brillouin zone of the (even,even,even) sublattice,
\bq
A^i=\left(\matrix{ 0&\alpha^i\cr\alpha^i&0\cr}\right)
\label{2.13}
\eq
and
\bq
\alpha^1=\left(\matrix{ 0&1\cr 1&0\cr}\right)
{}~~~
\alpha^2=\left(\matrix{ \sigma^1&0\cr 0&-\sigma^1\cr}\right)
{}~~~
\alpha^3=\left(\matrix{ \sigma^3&0\cr 0&-\sigma^3\cr}\right)
\label{2.14}
\eq
are a particular representation of the Dirac matrices.

In this representation the mass operator is
\bq
\sum_x (-1)^{\sum_{k=1}^{D}x_k}\psi^{\dagger}(x)\psi(x)=\int_{\tilde\Omega_B}
\frac{d^{3}k}{(2\pi)^3} \psi^{\dagger}(k)B\psi(k)
\label{2.15}
\eq
where
\bq
B=\left(\matrix{ 1&0\cr 0&-1\cr}\right)
\label{2.16}
\eq

The Fermion spectrum is
\bq
\omega(k)=\sqrt{\sum_{i=1}^3\sin^2k_i+m^2}
\label{fermspec}
\eq
and only the region $k_i\sim0$ is relevant to the continuum limit. We
have normalized $\psi(k)$ so that
\bq
\left\{\psi(x),\psi^{\dagger}(y)\right\}=\delta(x-y)
{}~~,~~
\left\{\psi(k),\psi^{\dagger}(l)\right\}=\delta(k-l)
\label{mscom}
\eq
If we define
\bq
\beta=\left(\matrix{\sigma^2&0\cr0&-\sigma^2}\right)
\eq
and the unitary matrix
\bq
M={1\over2}\left(\matrix{1-\beta&1+\beta\cr1+\beta&1-\beta\cr}\right)
\eq
and
\bq
\psi=M\psi'
\eq with
\bq
\psi'=(\psi_a,\psi_b)
\eq
the Hamiltonian is
\bq
H_f=\int_{\Omega_B}{d^{3}k}
\left(\psi_a^{\dagger},\psi_b^{\dagger}\right)
\left(\matrix{\alpha^i\sin k_i-\beta m&0\cr0&\alpha^i\sin k_i+\beta
m\cr}\right)\left(\matrix{\psi_a\cr\psi_b\cr}\right)
\eq
In the low momentum limit, $\sin k_i\sim k_i$, with Fermion density
1/2 per site so that the Fermi level is at the intersection point of
the positive and negative energy bands, we obtain 2 continuum Dirac
Fermions.

This describes two flavors of 4-component Dirac Fermions and the Dirac
masses for each component given by the staggered charge density have
opposite signs.  Thus the charge density breaks the discrete chiral
symmetry.  It also breaks a flavor symmetry which, in the absence of
mass, mixes the two continuum Fermions.

\subsection{\bf General Continuum Limit}

We shall now consider the continuum limit in a general number of
dimensions.  A formalism much like (but not exactly the same as) the
present one can be found in \cite{Kluberg}.

We shall begin with the Hamiltonian (\ref{2.7}),
\bq
H_f=-{i\over2}\sum_{[x,j]}(-1)^{\sum_{k=1}^{j-1}
x_k}\left(\psi^{\dagger}(x+\hat j)
\psi(x)-\psi^{\dagger}(x)\psi(x+\hat j)\right)
\nonumber
\eq
We consider an elementary hypercube of the lattice with sides of
length 1 and $2^d$ sites generated by taking a site all of whose
coordinates are even and adding to it the vectors
\bq
\vec\alpha = (\alpha_1,\ldots,\alpha_{D})
{}~~~~\alpha_i=0~{\rm OR~}1
\label{2.18}
\eq
We also decompose the lattice into $2^{D}$ sublattices generated by
taking a site of the elementary hypercube,
(even,even,...)+$\vec\alpha$ for some $\vec\alpha$ and translating it
by all even multiples of lattice unit vectors, $\hat i$.  We label the
Fermions which reside on the sublattice of each of the corners of the
elementary hypercube as $\psi_{\alpha_1,\dots,\alpha_{D}}(x)$.  In
momentum space the Hamiltonian is
\bq
H_f=\int_{\tilde\Omega_B}d^{D}k\sum_{i=1}^{D}
{}~\psi^{\dagger}_{\alpha_1,\ldots,\alpha_{D}}(k)
{}~\Gamma^i_{\alpha_1\ldots\alpha_{D}\beta_1\ldots\beta_{D}}\sin k_i
{}~\psi_{\beta_1\dots\beta_{D}}(k)
\label{2.19}
\eq
where $\tilde \Omega_B=\{ k_i:-\pi/2<k_i\leq\pi/2\}$ is the Brillouin zone of
the (even,even,$\ldots$) sublattice,
the momentum space Fermions have the anticommutator
\bq
\left\{\psi_{\alpha_1\dots\alpha_{D}}(k),
\psi^{\dagger}_{\beta_1\dots\beta_{D}}(k')\right\}=\left(\prod_{i=1}^{D}
\delta_{\alpha_i\beta_i}\right)\delta(k-k')
\eq
and the Dirac tensors are
\bq
\Gamma^i_{\alpha_1\dots\alpha_{D}\beta_1\ldots\beta_{D}}=\delta_{\alpha_1
\beta_1}\ldots
\delta_{\alpha_{i-1}\beta_{i-1}}\sigma^1_{\alpha_i\beta_i}
\delta_{\alpha_{i+1}\beta_{i+1}}\ldots
\delta_{\alpha_{D}\beta_{D}}(-1)^{\sum_{k=1}^{i-1}\alpha_k}
\label{2.20}
\eq
They obey the Clifford algebra
\bq
\Gamma^i\Gamma^j+\Gamma^j\Gamma^i= 2\delta^{ij}
\label{2.21}
\eq
The spectrum of the Dirac operator is
$\omega(k)=\pm\sqrt{\sum_{i=1}^D\sin^2k_i}$.  Note that, to set the
Fermi level of the Fermions at the degeneracy point where the two
branches of the spectrum meet, it is necessary that the Fermion states
are exactly half-filled.  This is also required for
charge--conjugation invariance, or particle--hole symmetry of the
vacuum state.

The staggered charge density operator (\ref{2.9}) is equivalent to a
mass operator where
\bq
\Sigma=\int_{\tilde\Omega_B}d^{D}k~
\psi^{\dagger}_{\alpha_1\ldots\alpha_{D}}(k)
\Gamma^0_{\alpha_1\ldots
\alpha_{D}\beta_1\ldots\beta_{D}}\psi_{\beta_1\ldots\beta_{D}}(k)
\label{2.22}
\eq
where
\bq
\Gamma^0_{\alpha_1\ldots\alpha_{D}\beta_1\ldots\beta_{D}}=
\delta_{\alpha_1\beta_1}\ldots\delta_{\alpha_{D}\beta_{D}}
(-1)^{\sum_{k=1}^{D}\alpha_k}
\label{2.23}
\eq
Here, $\Gamma^0$ satisfies the algebra
\bq
\Gamma^0\Gamma^i+\Gamma^i\Gamma^0=0
\label{2.24}
\eq
and
\bq
\Gamma^0\Gamma^0=1
\label{2.25}
\eq
Thus the spectrum of the operator $H_f+m\Sigma$ is
\bq
\omega(k)=\pm\sqrt{\sum_{i=1}^{D}\sin^2k_i +m^2}
\eq
which is the spectrum of a relativistic Fermion in the limit $k\sim
0$.

Here, we count the number of flavors of Fermions obtained in the
continuum limit by noting that (\ref{2.19}) describes a
$2^{D}$-component Fermion.  In $D$ dimensions the Dirac matrices are
$[(D+1)/2]$ dimensional, therefore the continuum limit of (\ref{2.19})
describes $2^{D}/2^{[(D+1)/2]}$ species of Dirac Fermions.  These are
tabulated up to dimension 4 in (\ref{matrix}).

\appende
\appa
\section{\bf Fermion Representation of SU($N$) Quantum Antiferromagnet}

The Hamiltonian for an U($N$) quantum antiferromagnet is
\bq
H_{\rm AFM}=\frac{g^2}{2}\sum_{<x,y>} J_{ab}(x)J_{ba}(y)
\label{AFM}
\eq
where $J_{ab}(x)$, $a,b=1,\cdots,N_{\tiny L}$, obey current algebra
relations associated with  the Lie algebra of U($N$),
\bq
\bigl[ J_{ab}(x),J_{cd}(y)\bigr]=\left(J_{ad}(x)\delta_{bc} -
J_{cb}(x)\delta_{ad}\right)\delta(x-y)
\label{algebra}
\eq
and where $<x,y>$ denotes the link connecting sites $x$ and $y$ on a
bipartite lattice.  For simplicity, we shall take the lattice to be
cubic.  Here, we have used a particular basis for the SU($N$) algebra
which can be conveniently represented by Fermion bilinear operators,
\bq
J_{ab}(x)=\psi^{a\dagger}(x)\psi^b(x)-\delta^{ab}/2
\label{generators}
\eq
The representation of the algebra on each site $x$ is fixed by
specifying the Fermion number of the states,
\bq
\rho(x)=\sum_a J_{aa}(x)
\eq
For example, the Fermion
vacuum state $\vert0>$ such that
\bq
\psi^a(x)\vert 0>=0~~~,~\forall a,x
\label{vac}
\eq
is the singlet state, the states with $m\leq N$ Fermions per site,
\bq
\prod_x \psi^{a_1\dagger}(x)\psi^{a_2\dagger}(x)\ldots\psi^{a_m\dagger}(x)
\vert 0>
\nonumber
\eq
corresponds to $\rho(x)=m-N/2$ for all $x$ and the irreducible
representation with the Young Tableau
\vskip 4.0truein
\noindent
{\bf Figure 3:}When there are $m$ Fermions per site, the representation of
$SU(N_L)$ has Young Tableau with one column of $m$ boxes.

\vskip 0.25truein

For each site $x$, $\rho(x)$ is the generator of the U(1) subgroup of
U($N$).  Using a basis $T^i=(T^i)^*$, $i=1,\ldots, N^2-1$,
of the Lie algebra of SU($N$) in the fundamental
representation normalized so that ${\rm tr}(T^iT^j)= T^i_{ab}T^j_{ba}
= \delta^{ij}/2$, and using
\bq
T^i_{ab}T^i_{cd}=
\frac{1}{2}\delta_{ad}\delta_{bc}-\frac{1}{2N}\delta_{ab}\delta_{cd}
\label{relSUN}
\eq
it is convenient to introduce
\bq
J^i(x) = \psi^{a\dagger}(x)T^i_{ab}\psi^a(x)
\eq
obeying current algebra of the Lie algebra of SU($N$), and to write
the Hamiltonian (\ref{AFM}) as
\bq
H_{\tiny AFM} = \frac{g^2}{N} \sum_{<x,y>}\rho(x)\rho(y) + H_{{\rm
SU}(N)}
\eq
with
\bq
H_{{\rm SU}(N)} = g^2\sum_{<x,y>} J^i(x)J^i(y)
\eq
is the Hamiltonian of an SU($N$) antiferromagnet.  From this it is
obvious that by fixing the $\rho(x)$, $H_{\tiny AFM}$ is reduced to an
SU($N$) antiferromagnet.

For example, the familiar $j=1/2$ SU(2) Heisenberg antiferromagnet is
obtained from (\ref{AFM}), $N=2$, by using the identity
\bq
\frac{\vec\sigma_{ab}}{2}\cdot\frac{\vec\sigma_{cd}}{2}=\frac{1}{ 2}
\delta_{ad}\delta_{bc}
-\frac{1}{2}\delta_{ab}\delta_{cd}
\label{identity}
\eq
corresponding to (\ref{relSUN}) for $N=2$.

Generally, when $N$ is even we will consider the representations where
$m=N/2$, so that the Fermion occupation of each site is $N/2$ and
$\rho(x)=0$.  When $N$ is odd we divide the lattice into two
sublattices such that the nearest neighbors of all sites of one
sublattice are in the other sublattice (when this is possible the
lattice is said to be bipartite).  When $N$ is odd, the representation
of SU($N$) has $(N+1)/2$ Fermions, i.e. $\rho(x)=1/2$, on the sites of
one sublattice and $(N-1)/2$ Fermions, i.e. $\rho(x)=-1/2$ on the
sites of the other sublattice.

\appende

\end{document}